# The First Stars May Shed Light on Dark Matter

*Abraham Loeb and Julián B. Muñoz*
*Harvard University, Cambridge, MA 02138, USA*

*Abstract:* Recent observations of hydrogen absorption that occurred when the first stars turned on may give insights into the nature of dark matter.

The nature of the dark matter is one of the longest-standing puzzles in cosmology. Astronomers have established that dark matter is the dominant constituent of matter in the Universe, but they are still in the dark about its identity. A possible clue may have been uncovered by recent observations of the cosmic dawn—the epoch when the first stars formed. Earlier this year, researchers reported a surprisingly strong absorption signal coming from gas activated by light from the first stars [1]. Now, a series of new papers [2-5] has explored what might be inferred about dark matter from this unexpected absorption. For example, the absorption could be explained by assuming that dark matter carries a small electric charge that allows it to interact weakly with ordinary matter. On the flip side, the absorption is inconsistent with certain models that predict dark matter should annihilate with itself. Regardless of the final interpretation, the cosmic dawn has clearly opened a new path toward resolving the dark matter puzzle.

For nearly a century [6], scientists have been studying dark matter through its gravitational effects on visible matter and radiation. Those observations have confirmed that dark matter is one of the primary constituents of the Universe. The measured anisotropies of the cosmic microwave background (CMB), for example, have shown that the overall density of dark matter is about six times that of ordinary (baryonic) matter. But the anisotropies are not the only aspect of the CMB that may contain information about the matter in the Universe. The CMB light carries an imprint of hydrogen gas that it encountered along its journey—a journey that started 400,000 years after the big bang. The imprinted signal is due to absorption of CMB photons with 21 cm wavelength, corresponding to the electronic transition in hydrogen's hyperfine levels (see Fig. 1). Because the universe is expanding, this absorption is redshifted to a longer wavelength, which depends on when the absorption occurred.

Early on during the so-called "dark ages," the absorption from hydrogen gas was minimal, as the populations in the hyperfine levels were in thermal equilibrium with the CMB. However, the absorption is expected to have increased dramatically at the start of the cosmic dawn—about a hundred million years after the big bang. At this time, the ultraviolet radiation from the first stars began exciting the hydrogen atoms, causing the hyperfine level populations to shift in such a way that they reflected the temperature of the gas [7]. As such, the strength of the 21-cm absorption from a particular epoch depends on the corresponding gas temperature, with more absorption occurring in colder gas. According to standard cosmological models, the gas reached its coldest temperature during the cosmic dawn (after which, the gas went through "reionization" and no longer absorbed light). However, interactions with dark matter could either cool [8] or heat [9] the hydrogen atoms from the cosmic dawn beyond the levels expected in Fig. 1.

**Figure 1:**

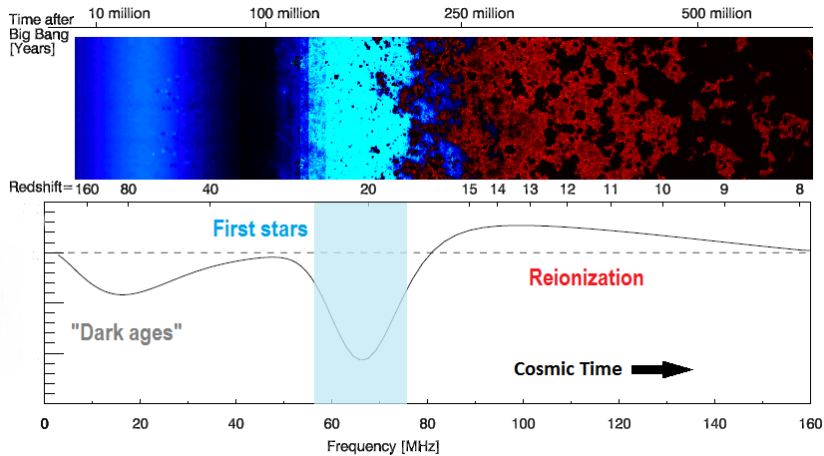

*Caption:* The 21 cm signal as a probe of the cosmic dawn. The top panel shows a simulation of hydrogen gas evolution in the Universe around the time when the first stars turned on. The colors indicate the brightness of the 21 cm line, with blue corresponding to absorption and red corresponding to emission. The bottom panel shows the prediction for the full-sky spectrum, with absorption occurring below the dashed line and emission above. Due to the cosmic redshift, each frequency maps to a specific time in the hydrogen gas evolution. Very early on, in the so-called "Dark ages", the absorption was relatively weak. But as stars turned on, their radiation helped to activate the 21-cm absorption, resulting in a big dip (marked by the blue bar). The recent observation from EDGES showed that this dip is deeper than predicted. At later times, the radiation from stars and black holes caused the gas to ionize, which eventually turned off the 21-cm signal.

*Credit:* Adapted from Pritchard, J., & Loeb, A. Reports in Progress in Physics, 75, 086901 (2012).

Measuring the 21-cm absorption is the goal of several new and upcoming experiments. One of these, EDGES (Experiment to Detect the Global Epoch of reionization Signal) [1], has offered the first sky-averaged spectrum at wavelengths corresponding to the redshifted 21-cm line. The data revealed an absorption dip at a wavelength of 380 cm, or equivalently, at a frequency of 78 MHz. The depth of this dip implies that roughly one in six CMB photons around this frequency were absorbed by intervening hydrogen. This depth is a factor of two larger than the standard prediction. It is tempting to associate this unexpected cooling with some non-standard interaction of the gas with cold dark matter particles, but one should proceed with caution. For instance, in Ref. [10] it was argued that a new fundamental force between dark matter and baryons could explain this signal. However, having a new force of this magnitude is ruled out because, for example, it would cause stars to cool faster than observed.

By contrast, we showed in [2] that the baryons can be cooled down if a fraction of the dark-matter particles are light in mass and have an electric charge of about one millionth that of the electron. Coulomb interactions would make these minicharged dark matter particles scatter off electrons and protons. Because the dark sector is colder than the baryonic sector, the net effect of these scatterings would be to cool the baryons. This scenario can explain the EDGES observations without introducing any new fundamental force. Asher Berlin from the SLAC National Accelerator Lab in California and colleagues have now considered a similar set-up and found results mirroring our own [3]. They also explored ways to constrain the abundance of minicharged dark matter by imagining that these particles can annihilate into neutrinos or into other types of dark matter.

Minicharged dark matter is not ruled out by current particle physics theory, but it nonetheless is an extraordinary claim that would require extraordinary evidence. Fortunately, the proposed interactions lead to a new testable prediction for the variation across the sky of the depth of the 21-cm absorption feature at 78 MHz [11]. Similar to how the intensity of the CMB depends on the direction you look, the 21-cm signal is expected to have spatial fluctuations based on the relative motion between dark matter and baryons [12]. Anastasia Fialkov from the Harvard-Smithsonian Center for Astrophysics in Massachusetts and her coworkers have now simulated multiple maps of the 21-cm signal, employing hundreds of different astrophysical models [4]. They show that adding a coupling between dark matter and ordinary matter significantly increases the 21-cm fluctuations relative to standard predictions. Therefore, future radio interferometers, such as LOFAR and HERA, should more easily detect the 21-cm fluctuations if the EDGES result is indeed caused by dark matter carrying a small electric charge.

Even if the EDGES signal is not the result of charged dark matter, a measurement of the gas temperature during cosmic dawn can still constrain the nature of dark matter in other ways. Guido D'Amico and his colleagues from CERN in Geneva now exploit this idea by studying how dark-matter annihilations, which are an outcome of many dark-matter models, could inject energy and heat the hydrogen [5]. They show that the EDGES observations—which imply colder-than-expected hydrogen—place the tightest constraints ever determined on dark matter annihilations. In particular, one can rule out typical dark-matter candidates if their masses are less than ten times the proton mass, eliminating the possibility of light, WIMP-like dark matter.

The above results showcase the promise of cosmic-dawn measurements for addressing the dark matter puzzle. Even feeble dark-matter interactions—which affect the hydrogen gas by either removing or depositing energy—can be constrained with 21-cm data. The 21-cm constraints complement those obtained from particle accelerators, which are more suited for probing stronger dark matter interactions.

To confirm the EDGES results, several additional experiments are underway, including the SARAS-2, LEDA, and PRIzM collaborations. The locations of these antennae in different hemispheres, as well as their different experimental designs, are essential for eliminating the possibility of instrumental or environmental contributions to the EDGES signal. Low-frequency interferometers, such as the previously mentioned LOFAR and HERA, will map the 21-cm fluctuations on the sky and verify the validity of the reported sky-averaged signal using an array of antennae instead of a single detector.

Studies of the cosmic dawn have a bright future ahead. When one of us published a textbook on this field five years ago, when data were scarce [13], he hoped that new physics unraveled by discoveries at 21 cm would require the book be revised. The reported EDGES signal brings us closer to fulfilling this hope.

This research is published in *Physical Review Letters*.

*About the Authors:*

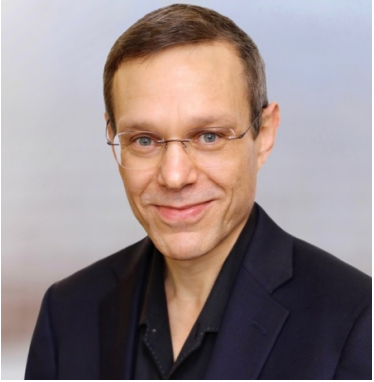

Abraham (Avi) Loeb is the Frank B. Baird, Jr., Professor of Science at Harvard University. Loeb published 4 books and over 600 papers on a wide range of topics, including black holes, the first stars, the search for extraterrestrial life and the future of the Universe. At Harvard, he serves as Chair of the *Department of Astronomy*, Founding Director of the *Black Hole Initiative (BHI)* and Director of the *Institute for Theory and Computation (ITC)*. He also chairs the *Board on Physics and Astronomy of the National Academies* which oversees the decadal surveys, serves as the Science Theory Director for all Initiatives of the *Breakthrough Prize Foundation*, as well as chair of the Advisory Committee for the *Breakthrough Starshot Initiative*. He is an elected fellow of the *American Academy of Arts & Sciences*, the *American Physical Society*, and the *International Academy of Astronautics*. In 2012, TIME magazine selected Loeb as one of the 25 most influential people in space.

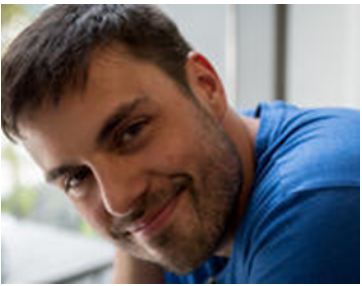

Julián B. Muñoz obtained his PhD from Johns Hopkins University in 2017, and is a postdoctoral fellow in the Jefferson Physical Laboratory, at Harvard University. He works in different aspects of theoretical cosmology, including dark-matter theories and large-scale structure.